\documentclass[aoas,preprint]{imsart}

\RequirePackage{amsthm,amsmath,amsfonts,amssymb}
\RequirePackage[authoryear]{natbib}
\RequirePackage[colorlinks,citecolor=blue,urlcolor=blue]{hyperref}
\RequirePackage{graphicx}
\usepackage{verbatim}
\usepackage{float}
\usepackage{enumitem}
\usepackage{multirow}

\usepackage{bm}

\newcommand{\bt}{\mbox{\boldmath$\beta$}}

\newcommand{\bx}{\mbox{\boldmath$x$}}
\newcommand{\X}{\mbox{\boldmath$X$}}

\usepackage{xcolor}

\newcommand{\PI}{\mbox{\boldmath$\pi$}}

\renewcommand{\eqref}[1]{(\ref{#1})}

\newcommand{\Q}{\mbox{\boldmath$\Theta$}}
\newcommand{\q}{\mbox{\boldmath$\theta$}}

\newcommand{\z}{\mbox{\boldmath$z$}}

\startlocaldefs
\theoremstyle{plain}

\theoremstyle{definition}

\endlocaldefs

\begin{document}

\begin{frontmatter}
\title{Tests of Covariate Effects and Model Selection in Finite Mixture Regression Model}
\runtitle{Tests and Model Selection in Gaussian Mixture Regression}

\begin{aug}
\author[A]{\fnms{Chong}~\snm{Gan} \ead[label=e1]{gan.c@queensu.ca}},
\author[B]{\fnms{Jacqueline}~\snm{May}\ead[label=e2]{jmay@uwaterloo.ca}}
\and
\author[C]{\fnms{Zeny}~\snm{Feng}\ead[label=e3]{zfeng@uoguelph.ca}}

\address[A]{Department of Public Health Sciences, Queen's University \printead[presep={ ,\ }]{e1}}

\address[B]{School of Pharmacy,  University of Waterloo 
\printead[presep={,\ }]{e2}}
\address[C]{Department of Mathematics and Statistics, University of Guelph \printead[presep={ ,\ }]{e3}}
\end{aug}

\begin{abstract}
Finite mixture regression models are widely used to capture the heterogeneous effects of covariates on a response variable. Our study, motivated by the analysis of Chiroptera, explores the dynamic effects of various environmental, biological, and human-related covariates on the evolutionary development of bat forearm using finite Gaussian mixture regression (GMR) models.  In this paper,  we address two key questions: (1) which covariates are associated with the variation in forearm length; and (2) whether the associated  covariates have the same effect across all subpopulations. To answer these questions,  we propose two stepwise procedures within the finite GMR framework based on the weighted significance test (WEST):  (1) a stepwise variable selection procedure to identify relevant influential covariates, and (2) a stepwise effect-type determination procedure to determine each relevant covariate as having either a homogeneous or heterogeneous effect across subpopulations.  Based on these procedures, we introduce a stepwise model selection procedure for  finite mixture regression, illustrated through finite GMR models.   Simulation studies are conducted to assess the performances of our proposed procedures. The proposed procedure will be applied to the Chiroptera dataset.
\end{abstract}


\end{frontmatter}

\section{Introduction}
Chiroptera (bats),  with over 1,400 species worldwide, are the second largest order of extant mammals (\cite{sotero2017chromosomal,wieringa2022comparing}). As the only mammals capable of true flight, they play important roles in ecosystem functions, including pollination, seed dispersal, and insect control (\cite{kunz2011ecosystem}). Moreover,  Chiroptera are highly sensitive to climate change and  often serve as a bioindicator for detecting environmental shifts (\cite{jones2009carpe}). Therefore,  it is crucial to understand the morphological response of Chiroptera to the changes of environment.  Forearm length,  measured as the distance from the elbow to the wrist,   is commonly used to assess their flight efficiency and overall adaptability (\cite{vannatta2022sexual}).   Hence,  it is of great importance to understand how environmental factors influence the morphological variability of bat forearms. 

Several studies have investigated the potential factors related to the evolutionary adjustments in bats' forearm length, yet findings remain inconclusive.  For example,  temperature may have an effect on bat's forearm development.  Bergmann's rule states that larger body size is advantageous for  effective thermoregulation at higher latitudes to withstand decreasing temperatures, suggesting that bats in colder regions may have longer forearms to retain heat  (\cite{bergmann1848verhaltnisse}). However, \cite{wood2023variability}  analysed data from 464 museum specimens of three bat species collected over 180 years and found that  warmer environments do not influence the forearm length of bats.   \cite{yue2020body} investigated the evolution of bat forearm length in response to increasing temperatures  in China over a study period of 65 years.  They  observed a significant increase in the forearm length of \textit{Hipposideros armiger}, while no change was detected in \textit{Rhinolophus leschenaultii}.   Geographical variation is another possible cause that has been empirically observed across bat species.    \cite{gohli2016interspecific} and \cite{de2014application} found that forearm length decreases in response to increasing latitude in \textit{Pteropodidae} and \textit{Carollia perspicillata}.  As latitude often serves as a proxy for temperature, these findings align with Allen’s rule,  which states that smaller appendages help animals conserve heat in colder environments at higher latitudes.  \cite{krystufek1993geographic} had further explained that this might be because bats in northern regions are born at a later time of a year  and enter hibernation earlier compared to those in southern regions .   These different latitudinal conditions also influence the availability of food resources.  Consequently, to adapt to their ecological requirements, bats develop forearm shapes and lengths that enhance their abilities to access food resources in their habitats (\cite{wood2023variability}).  Bat forearm length is also influenced by human activities. Increasing human population density alters land use and foraging habitats for bats. For example,  rapid urbanization in Italy after World War II led to a significant increase in street lighting.  \cite{tomassini2014cranial} found that street lights have become crucial foraging sites for bats.  As their diet shifted from small nocturnal dipterans to larger moths, these bats adapted by developing longer forearm lengths and larger cranial sizes.  According to studies from \cite{hughes2023big},  four out of five bat species have larger forearm in  urban areas due to longer prey availability seasons. Building on insights from the literature,  it appears that many factors have dynamic  effects on bat forearm evolutionary development and these effects vary among different bat species. Hence, a single‑population regression model may obscure important subpopulation‑specific relationships arising from unobserved heterogeneity.

In this study, we analyze a species‑level records for 589 bat species extracted from PanTHERIA (\cite{jones2009pantheria}), a comprehensive database containing biological and environmental information for thousands of mammal species. Forearm length is treated as the response, and the potential factors include population density, latitude, precipitation, evapotranspiration, temperature, area and body mass. These covariates summarize key biological and environmental influences that shape bat morphology and allow us to assess their association with forearm length and whether their effects are different across underlying unobserved subgroups.

Finite mixture regression (FMR) models have been widely used to model the heterogeneous covariate effects on the response and the underlying heterogeneous population structures. The probability distribution function of a response variable $y$ in a $G$-component FMR model can be expressed as
\begin{equation}
	f(y; \Q) = \sum_{g=1}^G  \pi_g f(y; \bt_g, \q_g).
\label{FiniteMixture}
\end{equation}
Here, $f(y; \bt_g, \q_g)$ is a distribution function characterized by parameters $\bt_g$ and $\q_g$,  where $\bt_g$ is a vector of intercept and  regression coefficients quantifying the strength of effects of covariates specific to the $g$th component and $\q_g$ represents other distribution parameter(s) characterized to the $g$th component.   The mixing proportions $\pi_g$ satisfy $0\leq\pi_g\leq 1$ for $g=1, ... , G$ with $\sum_{g=1}^G \pi_g=1$.  We let $\Q$ be a vector comprising all parameters.  In a finite mixture regression model,  each observation has a probability $\pi_g$ of belonging to component $g$,  and $\z_i=(z_{i1}, ..., z_{iG})^{'}$ is an unobserved vector of binary variables indicating the membership of $i$th observation.   Then,  in a finite Gaussian mixture regression (GMR) model, if a sample ($\bx_i, y_i$) belongs to component $g$,  $z_{ig}=1$ and $z_{ig'}=0$ for $\forall g'\neq g$ and the relationship between the mean of response $y_i$ and covariates $\bx_i$, is modelled as  $E(y_i|\bx_i,  z_{ig}=1) =  \bx_i \bt_g$. Hereafter, a simple term ``mixture" will be used for ``finite mixture".

In the most general FMR specification, all regression coefficients are component‑specific, so each  component has its own covariate–response relationship. At the other extreme, if the regression coefficients are constrained to be identical across components  (i.e. $\bt_1=...=\bt_G$), meaning that each covariate have the same effect on the response for all components, the FMR reduces to the standard generalized linear regression model (GLM) if  the distribution of the response is a member of the exponential family. However, allowing all covariate effects to be component‑specific can lead to unnecessarily complex models, whereas forcing all effects to be homogeneous may oversimplify the underlying structure. This motivates us to propose a more flexible mixture model that allows a subset of covariates  to have heterogeneous effects if their effects are supposed to be different, and others to have homogeneous effects across components if their effects are supposed to be the same. This idea is somewhat similar to the varying-coefficient models of \cite{hastie1993varying}, who generalize GLM by allowing regression coefficients to be either constant or to vary as a function of an observed effect-modifying variable.

Under this mixed effect-type FMR framework, our study aims to answer two key questions: (1) which covariates are significantly associated with forearm length development among bat species worldwide; and (2) for each relevant covariate, whether its effect is homogeneous or heterogeneous across underlying subpopulations.  We then seek to identify an ``optimal'' FMR model in terms of relevance and effect types, where relevant covariates are allowed to have different effect types (i.e., homogeneous or heterogeneous) across different components,  so as to best capture dynamic effects on bat forearm development.

In the previous study (\cite{gan2024tests}),  we found that the  Gaussian mixture regression model effectively clusters bat species regarding their forearm lengths of bats species into subpopulations where precipitation poses different effects on their forearm length development across different subpopulations.  Although  precipitation is one of the environmental variables that is widely regarded as having controversial effects on the forearm length development in literature,  it was the sole factor considered in the previous study. Given the mixed results in the literature,  it is plausible that forearm development  responds differently to multiple environmental factors across subpopulations. \cite{grun2008flexmix} present a FMR framewrok that  allows covariate effects to be specified as either homogeneous  or heterogeneous, but they do not provide statistical procedures for variable selection or for determining effect types from the data. 

To address these gaps, we propose forward and backward variable selection procedures for identifying relevant covariates. These procedures are based on testing the association between each covariate and the response using the weighted significance test (WEST). We then propose forward and backward  effect-type determination procedures to assess whether each selected covariate has a homogeneous or heterogeneous effect across components, again using WEST to conduct the corresponding tests. In \cite{gan2024tests}, the WEST procedure was developed for testing the overall effect and heterogeneous effect of a single covariate under a GMR framework in which the order of the mixture is unknown and the model includes only that single covariate. Here, we generalize WEST to accommodate multiple covariates, enabling overall effect and heterogeneous effect testing for each covariate within the nested GMR models. Following these sequential procedures, an optimal model in terms of  covariate relevance and effect type can be determined.

The paper is organized as follows: Section 2 introduces the hypothesis tests of covariate effects in nested GMR models.  The stepwise procedures for variable selection and effect-type determination are outlined in Section 3. The application of the  Chiroptera dataset is in Section 4 and the simulation study is present in Section 5.  Section 6 contains conclusions and discussions.
\section{Testing Covariate Effects in Nested Gaussian Mixture Regression Models}
\label{sec:methods}
Under the mixture regression framework, a covariate may or may not influence the response. When it does, the effect can be either homogeneous or heterogeneous across subpopulations.  Accordingly, the effects of a covariate on the response
can be classified into three types: (1) the covariate has no effect on the response,  (2) the covariate has the same effect on the response across all subpopulations,  and (3) the  covariate has different effects on the response across the subpopulations.  To determine which type of effect  a covariate has,   we conduct two hypothesis tests, (a) the null hypothesis that the regression coefficients  of the  covariate are zero across all subpopulations,  against the alternative hypothesis that at least one of the regression coefficients is non-zero, and (b) the null hypothesis that  the regression coefficients  are equal and non-zero across all subpopulations,  against the alternative that the regression coefficients in at least one subpopulation differ from those in the others.  We refer to  these as  \textit{overall effect test} and \textit{heterogeneous effect test}, respectively.  These two tests can be applied separately or sequentially depending on the application.  When only a single covariate (i.e. $\bx_1$) is considered, details of overall effect test and heterogeneous effect test under GMR model can be found in  \cite{gan2024tests}. The  covariate-effect tests  in a nested GMR model with multiple covariates are presented below.

\subsection{Overall effect test}
The overall effect test is used to test for the significance of an additional covariate given that there are other covariates in the  model.  Note that for all covariates we consider, we do not restrict any of them to have homogeneous effects. Suppose there are $k-1$ covariates
in a GMR model with $G$ components, where $G$ is known or pre-specified.  The model can be represented as
\begin{equation}
y_i=\beta_{0g}+\beta_{1g}x_{i1}+...+\beta_{k-1, g} x_{i, k-1}+\epsilon_{ig}, \ \ \epsilon_{ig}  \overset{\mathrm{iid}}{\sim} N(0,\sigma_g^2),
\label{gmrj-1x}
\end{equation}
for $i=1,...,n$ and $g=1,...,G$.  Now when the   $k$th covariate, $\bx_k$, is added to the model, the model is  represented as 
\begin{equation}
y_i=\beta_{0g}+\beta_{1g}x_{i1}+...+\beta_{k-1, g} x_{i, k-1}+\beta_{kg}x_{ik}+\epsilon_{ig}, \ \ \epsilon_{ig}  \overset{\mathrm{iid}}{\sim} N(0,\sigma_g^2).
\label{gmrpx}
\end{equation}
An overall effect test for  $\bx_k$ as
\begin{equation}
\begin{aligned}
\label{eq:hypothesis1-px}
H_{0}&:{\beta_{k1}}=\dots={\beta_{kg}}=\dots={\beta_{kG}}={0}\\H_{a}&: \exists{g}\in{G} \mbox{ such that }\beta_{kg}\neq 0
\end{aligned}
\end{equation}
can be used to determine whether $\bx_k$ should be included in the model. The null hypothesis corresponds to a nested model with $k-1$ covariates as in (\ref{gmrj-1x}) and  the alternative hypothesis corresponds to a  full model with $k$ covariates as in (\ref{gmrpx}).  When $k=1$, the model under $H_0$ is reduced to a Gaussian mixture (GM) model without covariates and the model under the $H_a$ is a GMR model with one covariate only.  

When the order $G$ is known or pre-specified,   the likelihood ratio test (LRT) can be employed.  The statistic for the LRT is expressed as
 \begin{equation}
 \text{LRT} =-2(l(\hat{\Q}_{\text{GMR-nested}})-l(\hat{\Q}_{\text{GMR}})).
\label{eq:LRTov}
\end{equation}
The LRT statistic asymptotically conforms to a $\chi^2_G$ distribution,  where $\hat{\Q}_{\text{GMR-nested}}$ and $\hat{\Q}_{\text{GMR}}$ represent the maximum likelihood estimates (MLEs) of $\Q$ for the nested model  in (\ref{gmrj-1x})  and the  full model in (\ref{gmrpx}), respectively,  obtained using EM algorithms.  $l(\hat{\Q}_{\text{GMR-nested}})$ and $l(\hat{\Q}_{\text{GMR}})$ are their corresponding maximum log likelihood values respectively.

\subsection{Heterogeneous effect test}
\label{method:het}
When a covariate is found to have a significant effect on the response based on the overall effect test, the heterogeneous effect test can then be  determined whether   the covariate's effect is  homogeneous or heterogeneous  across all subpopulations. Suppose that $k$ covariates have been found to have significant effects on the response. Among these covariates, let $q$ of them ($0<q<k-1$) have heterogeneous effects, and the remaining  $k-1-q$ have homogeneous effects. Accordingly, the regression coefficients for the $q$ covariates are left unconstrained to be equal across components, whereas those for the $k-1-q$ covariates are constrained to be equal (but nonzero) across components. We now focus on testing whether the $k$th covariate has a homogeneous effect or heterogeneous effects. The heterogeneous effect test for $\bx_k$ is formulated as follows:
\begin{equation}
\begin{aligned}
\label{eq:hypothesis2-px-j}
H_{0}&:{\beta_{k1}}=\dots={\beta_{kg}}=\dots={\beta_{kG}}={\beta_k}\\
H_{a}&: \exists{g \neq g' }\in{G} \mbox{ such that }\beta_{kg}\neq \beta_{kg'},
\end{aligned}
\end{equation}
for $\beta_k\neq0$. Under the null hypothesis, the coefficient vector
$\bt_k$ is constrained to be the same across all components, and the corresponding model becomes
\begin{equation}
y_i=\beta_{0g}+\beta_{1g}x_{i1}+...+\beta_{qg} x_{iq}+\beta_{q+1} x_{i, q+1}+...+\beta_{k-1} x_{i, k-1}+\beta_{k} x_{ik}+\epsilon_{ig}, \ \ \epsilon_{ig}  \overset{\mathrm{iid}}{\sim} N(0,\sigma_g^2).
\label{GhomoH0}
\end{equation}
Under the alternative hypothesis $H_a$,  the coefficients $\bt_k$ are unconstrained, and the resulting model can be written as
\begin{equation}
y_i=\beta_{0g}+\beta_{1g}x_{i1}+...+\beta_{qg} x_{iq}+\beta_{q+1} x_{i, q+1}+...+\beta_{k-1} x_{i, k-1}+\beta_{kg} x_{ik}+\epsilon_{ig}, \ \ \epsilon_{ig}  \overset{\mathrm{iid}}{\sim} N(0,\sigma_g^2).
\label{GhomoHa}
\end{equation}
The LRT is applicable for this heterogeneous effect test.   The LRT statistic of the heterogeneous effect of $\bx_k$  is given as
\begin{equation}
\begin{aligned}
\text{LRT}
          =-2(l(\hat{\Q}_{H_0})-l(\hat{\Q}_{H_a})).
\label{eq:LRT_he2}
\end{aligned}
\end{equation}
The LRT statistic follows $\chi^{2}_{G-1}$ asymptotically, where $\hat{\Q}_{H_0}$ and $\hat{\Q}_{H_a}$ represent the MLEs for the model under the $H_0$ and $H_a$,  with $l(\hat{\Q}_{H_0})$ and $l(\hat{\Q}_{H_0})$ being their respective maximum log likelihood values.  

The hypothesis test in (\ref{eq:hypothesis2-px-j}) can be used to develop the stepwise procedures to determine the effect type of each relevant covariate.

\subsection{WEST Procedure for overall and heterogeneous effect tests}
\label{sec:o&h}
The LRTs for covariate effects,  as discussed in the previous section,  assume that the order of a mixture model is predetermined or known.  Yet, in real-world scenarios,  the order of a mixture population often remains unknown.  The optimal order $G$ selected under $H_0$ and $H_a$ can be different.  The uncertainty in determining the order $G$ can be complicated. To avoid the issues arising from order specification that may bias hypothesis test results,  WEST procedure that aggregates the statistical evidences on models with a range of orders was proposed in \citet{gan2024tests}.  In this paper,  the WEST procedure will be used for the hypothesis tests in (\ref{eq:hypothesis1-px}) and (\ref{eq:hypothesis2-px-j}). 

Let $O=\{G_1, ..., G_J\}$ be a range of $J$ candidate orders with $G_1$ and $G_J$ being the minimum and maximum order respectively.  Let $M_j$ denote the model of order $G_j$  under $H_0$ of a test.  For a pre-specified significance level $\alpha$, WEST procedure  is outlined as follows:

 \begin{enumerate}[leftmargin=0.55in, topsep=0pt,itemsep=0ex,partopsep=1ex,parsep=1ex]
\item[step 1:]  Fit models under $H_0$ and $H_a$ for each order  in $O$.
\item[step 2:]  Compute the observed LRT statistic, $LRT_{obs}$,  for each order in $O$. 
\item[step 3:]  Compute the p-value for each order in $O$ and denote the p-value as $\text{p-val}_j$ for $j=1, ..., J$.
 \[\text{p-val}_j = P( \mbox{LRT}_j \geq \mbox{LRT}_{obs_j}|M_j)\]
where under the null model $M_j$,  $\mbox{LRT}_j$ follows $\chi^2_{\text{df}_j, \alpha}$ asymptotically with $\text{df}_j$ being the difference in the number of parameters between the models  under $H_a$ and $H_0$ at the order of $G_j$.
\item[step 4:]  Compute the weighted significance level $p_0$ where 
\[p_0=\sum_{j=1}^J (\text{p-val}_j \cdot \Pi_j)\]
and $\Pi_j \approx \frac{\exp(-\frac{1}{2}\text{BIC}(M_j))}{\sum_{j=1}^J\exp(-\frac{1}{2}\text{BIC}(M_j))}$.
\item[step 5:]  Reject the $H_0$ if $p_0$ is smaller than prespecified  $\alpha$, or otherwise do not reject
$H_0$.
\end{enumerate}

\section{Stepwise Procedures for Variable Selection and Effect-Type Determination}
\label{method:TPMS}
In real applications, when we have a set of covariates, say $p$ covariates, their effect types on the response are usually unknown.  It is important to determine the effect type for each covariate.   To do so, we begin by applying a stepwise variable selection procedure based on the overall effect tests to identify a subset of relevant covariates, say $p^*$ ($\leq p$) covariates, that influence the  response.  Once the relevant covariates are selected, we proceed with a stepwise effect-type determination procedure based on heterogeneous effect test to determine for each of the $p^*$ covariates whether its effect is  homogeneous or heterogeneous.  Stepwise procedures for variable selection and effect-type determination are introduced below.  In both procedures, we consider the forward and backward options.

\subsection{Stepwise variable selection procedures}
\subsubsection{Forward variable selection}
Let $\bx_1, \bx_2, ... , \bx_p$ denote the $p$ covariates considered in the GMR model.  The proposed forward variable selection procedure is outlined as follows:

 \begin{enumerate}[leftmargin=0.55in, topsep=0pt,itemsep=0ex,partopsep=1ex,parsep=1ex]
 
 	\item[step 1:]  For each order in $O$, fit a GMR model without covariates (i.e. GM model) as:
\begin{equation}
	y_i=\beta_{0g}+\epsilon_{ig}, \ \ \epsilon_{ig}  \overset{\mathrm{iid}}{\sim} N(0,\sigma_{g}^2).
	\label{full0}
	\end{equation}
for $g=1, ... G$. 

	\item[step 2:]  Perform WEST procedure to test
	\[\begin{aligned}
		H_{0}&:{\beta_{j1}}=\dots={\beta_{jg}}=\dots={\beta_{jG}}={0}\\H_{a}&: \exists{g}\in{G} \mbox{ such that }\beta_{jg}\neq 0
	\end{aligned}\]
	for $j=1,...,p$, where $H_0$ corresponds to a GM model without covariates as in (\ref{full0}), and $H_a$ corresponds to a GMR model with a single covariate $\bx_j$. 
	If $p_0 > \alpha$ for all covariates,  suggesting no covariate is associated with the response,  stop the selection procedure, and no covariates are included in the model. Otherwise,  select the covariate that has the smallest $p_0$ value and add it to the model. Once selected, this covariate remains in the model and is removed from the candidate covariate set for the later tests..
	
\noindent Repeat Step 2 for the remaining candidate covariates, conditional on the covariates already included in the model, until no additional covariate is selected. This  completes the forward selection procedure. 

\end{enumerate}

\subsubsection{Backward variable elimination}
Suppose we consider $p$  covariates: $\bx_1, \bx_2, ... , \bx_p$, the backward elimination procedure is outlined as follow: 

 \begin{enumerate}[leftmargin=0.55in, topsep=0pt,itemsep=0ex,partopsep=1ex,parsep=1ex]
	\item[step 1:]  For each order in $O$, fit a GMR model that includes all variables as:
\begin{equation}
	y_i=\beta_{0g}+\beta_{1g}x_{i1}+...+\beta_{kg} x_{ik}+...+\beta_{pg}x_{ip}+\epsilon_{ig}, \ \ \epsilon_{ig}  \overset{\mathrm{iid}}{\sim} N(0,\sigma_{g}^2).
	\label{full}
	\end{equation}
for $g=1, ... G$. 

	\item[step 2:]  Perform WEST procedure to test
	\[\begin{aligned}
		H_{0}&:{\beta_{j1}}=\dots={\beta_{jg}}=\dots={\beta_{jG}}={0}\\H_{a}&: \exists{g}\in{G} \mbox{ such that }\beta_{jg}\neq 0
	\end{aligned}\]
	where $H_0$ corresponds to a GMR model including all covariates except $\bx_j$, for $j=1, ..., p$, and $H_a$ corresponds to the GMR model with all covariates as in (\ref{full}).  If $p_0 < \alpha$ for all covariates, stop the elimination procedure as all covariates are  significant and so no covariate will be removed from model (\ref{full}).  Otherwise,  remove the covariate that has the largest $p_0$ value and reset the remaining covariates as $\bx_1, \bx_2, ... , \bx_{p-1}$ for a reduced candidate set.
	
	\noindent Repeat step 2 for the reduced candidate set until no further covariates can be removed. This completes the backward elimination procedure.

	\end{enumerate}
	\noindent If one is interested only in variable selection, the stepwise variable selection procedure can be used alone without proceeding to the stepwise effect‑type determination. With the selected covariates, we can select the optimal order of the GMR with the one having the lowest BIC among all candidate orders in $O$.
\subsection{Stepwise effect-type determination procedures}
\subsubsection{Forward effect-type determination procedure}
Following the stepwise variable selection procedure,  suppose $p^*$
  covariates are selected as relevant to the variation in the response.  For notational simplicity, we relabel these selected covariates  as $\bx_1,  \bx_2, ...\bx_{p^*}$. Building on this selection,  we propose a forward procedure to determine the effect type of either homogeneous or heterogeneous for each of the selected covariates, as described below.

 \begin{enumerate}[leftmargin=0.55in, topsep=0pt,itemsep=0ex,partopsep=1ex,parsep=1ex]
 \item[step 1:]  For each order in $O$,  fit a GMR model that includes all covariates $\bx_1,  \bx_2, ...\bx_{p^*}$, where all covariates have homogeneous effects across all subpopulations, as follows:
\begin{equation}
	y_i=\beta_{0g}+\beta_{1}x_{i1}+...+\beta_{k} x_{ik}+...+\beta_{p^*}x_{ip^*}+\epsilon_{ig}, \ \ \epsilon_{ig}  \overset{\mathrm{iid}}{\sim} N(0,\sigma_g^2).
	\label{full2}
	\end{equation}
for $g=1, ... G$. 

	\item[step 2:]  Perform WEST procedure to test
	\[\begin{aligned}
		H_{0}&:{\beta_{j1}}=\dots={\beta_{jg}}=\dots={\beta_{jG}}={\beta_j}\\H_{a}&: \exists{g}\in{G} \mbox{ such that }\beta_{jg}\neq \beta_{jg'}
	\end{aligned}\]
for $\beta_j \neq 0$ and $j=1, ..., p^*$.  Here, $H_0$ corresponds to a GMR model in which all covariates have homogeneous effects on the response $y$, as in (\ref{full2}) and $H_a$ corresponds to   a GMR model where covariate 
$\bx_j$ has a heterogeneous effect, while all other covariates have homogeneous effects.   If $p_0 > \alpha$ for all covariates,  suggesting all covariates have homogeneous effects on the response, stop the procedure.  Otherwise, set $\beta_{j'g}$ to be unrestricted (i.e., treat the $j'$th covariate as having a heterogeneous effect) if its $p_0 < \alpha$ and is the smallest among $p_0$ values for all covariates.  

\noindent Repeat step 2 for all $j \neq j'$, conditional on $\bx_{j'}$ having a heterogeneous effect, set the coefficient of the most significant covariate to be unrestricted, and continue until no further covariates are found to have significant heterogeneous effects. This completes the forward effect-type determination procedure. 
\end{enumerate}

\subsubsection{Backward effect-type determination procedure}
The backward procedure to determine the effect type of $p^*$ selected covariates, $\bx_1,  \bx_2, ...\bx_{p^*}$, is proposed as follow:

 \begin{enumerate}[leftmargin=0.55in, topsep=0pt,itemsep=0ex,partopsep=1ex,parsep=1ex]
	\item[step 1:]  For each order in $O$, fit a GMR model that includes all $p^*$ covariates with heterogeneous effects  as:
\begin{equation}
	y_i=\beta_{0g}+\beta_{1g}x_{i1}+...+\beta_{kg} x_{ik}+...+\beta_{p^*g}x_{ip^*}+\epsilon_{ig}, \ \ \epsilon_{ig}  \overset{\mathrm{iid}}{\sim} N(0,\sigma_g^2).
	\label{full3}
	\end{equation}
for $g=1, ... G$. 

\item[step 2:]  Perform WEST procedure to test
	\[\begin{aligned}
		H_{0}&:{\beta_{j1}}=\dots={\beta_{jg}}=\dots={\beta_{jG}}={\beta_j}\\H_{a}&: \exists{g}\in{G} \mbox{ such that }\beta_{jg}\neq \beta_{jg'}
	\end{aligned}\]
for $\beta_j \neq 0$ and $j=1, ..., p^*$.  If $p_0 < \alpha$ for all covariates,  stop the procedure as there are significant evidence to suggest that none of the covariates have a homogeneous effect on the response across subpopulations.   Otherwise,  set $\beta_{j'g}=\beta_{j'}$ to be restricted (i.e., treat the $j'$th covariate as having a homogeneous effect) if its $p_0 > \alpha$ and is the largest among $p_0$ values for all covariates.  \\[2mm]
  \noindent Repeat step 2 for all $j \neq j'$, conditional on $x_{j'}$ having
homogeneous effects, set the coefficient of the most insignificant covariate to be restricted, and continue until no further covariates are found to have insignificant heterogeneous effects.  This completes the backward effect-type determination procedure. 

	\end{enumerate}
	
Fit the GMR model of each order in $O$ by including $p^*$ covariates with their effects types being specified according to the results of the effect-type determination procedure.  The optimal order is selected as the one with the lowest BIC value among all order candidates.

\section{Application to Chiroptera data}
\subsection{Data Description}
The data of Chiroptera  were sourced from the PanTHERIA database, which is accessible through Ecological Archives      (\cite{jones2009pantheria}).  We extracted species in the order of Chiroptera from PanTHERIA,  creating a dataset containing 589 bat species. This dataset comprises nine variables:   the forearm length of adult bats measured in mm from elbow to wrist (Forearm Length),   the average monthly precipitation in mm (Precipitation), the average monthly temperature in 0.1 \textdegree{C} (Temperature),  mean monthly actual evapotranspiration rate (AET) in their habitats and activity areas, median latitudinal extent of the bat species' range (Latitude),  total extent of the bat species' range using a global equal-area projection (Area), body mass in gram (Body Mass),  human population density (Population Density) around the bats' habitats and activity areas,  and the trophic level (Trophic).   The forearm length is a crucial metric for studying bats' evolutionary adaptations,  and will serve as the response in our study, where other variables excluding the tropic level are potential covariates to be determined for their inclusion using in the mixture regression model.  Trophic level classifies a species according to its dietary patterns (herbivore, carnivore or omnivore).  Collecting accurate trophic level data for bats is challenging due to the diversity of dietary habits, wide geographical distribution and seasonal dietary changes,  requiring extensive fieldwork and analysis  (\cite{o2022spatial}).  Out of 589 species in our data set, only 358 have information of their trophic levels.   Therefore, we exclude the trophic level variable in our variable selection process as including this variable will greatly reduce the sample size.  Instead, we use the trophic level as a reference to see how dietary patterns relate to the clustering result of bat species with respect to their forearm lengths and the relationship between their forearm lengths and other variables.

\subsection{Results of sequential stepwise procedures}
Both forward and backward stepwise procedures are applied to the Chiroptera data set for selecting relevant covariates  that affect the  forearm length and for determining the effect type of the selected covariates.  Before model fitting,  transformation is required for some covariates.   Body Mass,   Population Density and Area show exponential growth or decay and contain very large outliers.  A log transformation is used to  reduce the skewness and scale down extreme values.  Since Population Density contains zero value,  a transformation of log(Population Density + 0.01)  is applied.  Latitude is transformed by taking its absolute value to adjust for its symmetry about the equator.  To determine the maximum order $G_J$  in the candidate set $O$ for the stepwise variable selection procedure,  we let $G_J$ start from 5 (under the $H_0$) and compute the approximate posterior probability $\Pi_ {G_J}$ for model $M_{G_J}$, as described in step 4 of the WEST procedure.  
The value of $G_J$ is incremented by one at each iteration until $\Pi_ {G_J}<10^{-4}$.  This process yields different maximum orders for the two procedures: $G_J=9$ for the forward procedure and $G_J=7$ for the backward procedure.  Accordingly, the candidate sets of component orders used in model selection are $O=\{2, ... , 9\}$ for both the forward procedure and the backward procedure.

Table \ref{Table:chiFS} presents step-by-step results of variable selection among 7 covariates  using the forward variable selection procedure.   For example, at the $k$th step for $k=1, ..., 7$,  $x_k$ represents the addition covariate from the remaining candidate set of $x_1, ..., x_{p-k+1}$ to be tested for their association with the response. The row of $p_0$ lists the corresponding weighted significance level for each  covariate.  The covariate with lowest significant $p_0$ (bold font) will be added to the model and involved in both $H_0$ and $H_a$ models in the subsequent step.  In the first step, among all covariates,  $\log$(Body Mass) is the first covariate added to the GMR model with the smallest $p_0$ value ($9.69*10^{-169}$),  followed by $\log$(Population Density) ($p_0=1.01*10^{-5}$) in step 2. Then abs(Latitude) ($p_0=5.94*10^{-4}$),  $\log$(Area) ($p_0=1.66*10^{-2}$),  AET ($p_0=1.25*10^{-2}$), Precipitation ($p_0=5.42*10^{-3}$) and Temperature ($p_0=3.27*10^{-3}$).  As a result, all covariates are selected. 


\begin{table}[h!] 
\centering 
  \caption {Forward variable selection for the Chiroptera data with all potential factors.  } 
  \label{Table:chiFS}
      \resizebox{\textwidth}{!}{
\begin{tabular}{cl  l l l l l l }  
\\[1mm]\hline 
\hline\\[-3mm]
\multicolumn{8}{l}{1-Covariate GMR Model ($Y\sim X_1$)}\\[1mm]
\hline\\[-3mm]
$X_1$&Precipitation&Temperature& abs(Latitude)&$\log$(Population Density)&$\log$(Area)&$\log$(Body Mass)&AET\\[1mm]
$p_0$&$1.69*10^{-4}$ & $2.38*10^{-3}$  &
$1.05*10^{-4}$ & $1.52*10^{-1}$& $1.49*10^{-1}$ & \bm{$9.69*10^{-169}$} & $6.99*10^{-2}$\\[1mm]
\hline\\[-3mm]
\multicolumn{8}{l}{2-Covariate GMR Model ($Y\sim \log(\text{Body Mass}) + X_2$)}\\[1mm]
\hline\\[-3mm]
$X_2$&Precipitation&Temperature&abs(Latitude)&$\log$(Population Density)&$\log$(Area)& AET&\\[1mm]
$p_0$&$7.14*10^{-2}$ & $4.68*10^{-4}$  &
$4.23*10^{-3}$ & \bm{$1.01*10^{-5}$}& $8.41*10^{-2}$ & $1.26*10^{-1}$ &\\[1mm]
\hline\\[-3mm]
\multicolumn{8}{l}{3-Covariate GMR Model ($Y\sim \log(\text{Body Mass}) +  \log(\text{Population Density})+X_3$)}\\[1mm]
\hline\\[-3mm]
$X_3$&Precipitation&Temperature&abs(Latitude)&$\log$(Area)& AET&&\\[1mm]
$p_0$&$5.44*10^{-2}$ & $1.41*10^{-3}$ & \bm{$5.94*10^{-4}$} &
$7.38*10^{-2}$ & $6.00*10^{-2}$ &&\\[1mm]
\hline\\[-3mm]
\multicolumn{8}{l}{4-Covariate GMR Model ($Y\sim \log(\text{Body Mass}) +  \log(\text{Population Density})+ \text{abs(Latitude)} +X_4$)}\\[1mm]
\hline\\[-3mm]
$X_4$&Precipitation&Temperature&$\log$(Area)& AET&&&\\[1mm]
$p_0$&$7.88*10^{-2}$ & $1.94*10^{-1}$ & \bm{$1.66*10^{-2}$} & $2.53*10^{-1}$ &&&\\[1mm]
\hline\\[-3mm]
\multicolumn{8}{l}{5-Covariate GMR Model ($Y\sim \log(\text{Body Mass}) +  \log(\text{Population Density})+\text{abs(Latitude)} + \log(\text{Area)} +X_5$)}\\[1mm]
\hline\\[-3mm]
$X_5$&Precipitation&Temperature& AET&&&&\\[1mm]
$p_0$&$3.27*10^{-2}$ & $2.32*10^{-1}$ & \bm{$1.25*10^{-2}$}&&&&\\[1mm]
\hline\\[-3mm]
\multicolumn{8}{l}{6-Covariate GMR Model ($Y\sim \log(\text{Body Mass}) +  \log(\text{Population Density})+\text{abs(Latitude)}+ \log(\text{Area)} + \text{AET}+ X_6$)}\\[1mm]
\hline\\[-3mm]
$X_6$&Precipitation&Temperature& &&&&\\[1mm]
$p_0$&\bm{$5.42*10^{-3}$} & $1.42*10^{-2}$&&&&&\\[1mm]
\hline\\[-3mm]
\multicolumn{8}{l}{7-Covariate GMR Model ($Y\sim \log(\text{Body Mass}) +  \log(\text{Population Density})+ \text{abs(Latitude)} + \log(\text{Area)} + \text{AET}+ \text{Precipitation} + X_7$)}\\[1mm]
\hline\\[-3mm]
$X_7$&Temperature&& &&&&\\[1mm]
$p_0$&$\bm{7.49*10^{-3}}$&&&&&&\\[1mm]
\hline
\hline\\[-3mm]
Preferred model&\multicolumn{7}{l}{ 4-component GMR model with 7 covariates: $\log(\text{Body Mass}),  \log(\text{Population Density}), \text{abs(Latitude)}, \log(\text{Area)}, \text{AET}, \text{Precipitation}, \text{Temperature}$}\\[1mm]
\hline
\hline
\end{tabular}}
\end{table} 

\begin{table}[h!] 
\centering 
  \caption {Backward elimination procedure for selecting variables in fitting GMR model for the Chiroptera data.  } 
  \label{backwardr1}
      \resizebox{\textwidth}{!}{
\begin{tabular}{c l  l l l l l l }  
\\\hline 
\hline\\[-3mm]
\multicolumn{8}{l}{6-Covariate GMR Model ($Y\sim X_{-1}$) ($X_{-1}$: exclude $X_1$)}\\[1mm]
\hline \\[-3mm]
$X_1$&Precipitation&Temperature& abs(Latitude)&$\log$(Population Density)&$\log$(Area)&$\log$(Body Mass)& AET\\[1mm]
$p_0$& $1.30*10^{-3}$ & $7.49*10^{-3}$ & 
\bm{$8.49*10^{-1}}$ & $1.75*10^{-5}$ &$2.70*10^{-3}$ & $1.08*10^{-178}$ &
$7.58*10^{-4}$ \\

\hline\\[-3mm]
\multicolumn{8}{l}{5-Covariate GMR Model ($Y\sim X_{-1,-2}$) ($X_{-1,-2}$: exclude abs(Latitude) and $X_2$ )}\\[1mm]
\hline\\[-3mm]
$X_2$&Precipitation&Temperature&$\log$(Population Density)&$\log$(Area)&$\log$(Body Mass)& AET\\[1mm]
$p_0$& $1.89*10^{-4}$ & $6.25*10^{-4}$ & $2.77*10^{-5}$ &
$4.45*10^{-2}$ & $1.35*10^{-186}$ & $8.09*10^{-5}$ \\

\hline
\hline\\[-3mm]
Preferred model&\multicolumn{7}{l}{ 4-component GMR model with 6 covariates: $\log(\text{Body Mass}),  \log(\text{Population Density}), \log(\text{Area)}, \text{AET}, \text{Precipitation}, \text{Temperature}$}\\[1mm]
\hline
\hline\\[-3mm]
\end{tabular}}
\end{table} 

Table \ref{backwardr1} summarizes results of variable selection using the backward  elimination procedure. In the first step, the abs(Latitude) is found to be nonsignificant ($p_0=0.849>0.05$) at the 5\% significance level, so abs(Latitude) is removed from the model. The procedure stops at step 2 because all six remaining covariates have $p_0<0.05$.  

The forward and backward procedures yield different sets of relevant covariates: the forward procedure includes abs(Latitude),  while the backward procedure excludes it.  For each procedure,  among all orders in their respective sets $O$,  we find the GMR model that achieves the lowest BIC using the selected covariates. For example, under the forward procedure, the GMR model including all seven covariates has the lowest BIC at $G=4$. The 4-component GMR with 6 covariates are preferred using backward procedure.   Comparing the BIC values of the two preferred models,  the model without abs(Latitude) has a lower BIC (4183.79) than the model including abs(Latitude) (4209.28), suggesting that the model without abs(Latitude) is more preferable. Therefore, abs(Latitude) is excluded from the final set of relevant covariates. The absolute value of latitude (abs(Latitude)) measures the distance from the equator and correlates with temperature and the average evapotranspiration rate (AET). It is often used as a proxy for temperature when temperature data is unavailable. When both temperature and AET are included in the model, it is reasonable that abs(Latitude) is no longer significant when testing its overall effects.  Figure S1 in Section 1 in the Supplemental Material present the visualization of clustering analysis results from the two preferred models using covariates log(Body Mass) and log(Population Density). There are only minor differences in the membership assignments of bat species. 
	
We now apply the stepwise effect-type determination procedures to assess whether each of the six selected covariates has a homogeneous or heterogeneous effect on the forearm development.	To do so, we still use the same order list as we used in the variable selection procedure as changing the effect type of each selected variable might change the fitness of models.  The results of forward effect-type determination procedure are shown in Table \ref{ForwardDET}.  In the first step, the smallest $p_0$ value for the log(Body Mass)  ($9.69\times 10^{-169}$) indicates a strongest evidence that the log(Body Mass) has heterogeneous effects on forearm lengths of bat species conditional on all other covariates having homogeneous effects, so log(Body Mass) is the first covariate to switch from having a homogeneous effect to a heterogeneous effect, then followed by $\log$(Population Density) ($p_0=3.52*10^{-4}$), and Precipitation ($p_0=4.30*10^{-5}$).  The remaining covariates, log(Area), AET and Temperature,  have $p_0>0.05$ at step 4,  suggesting that their effects on forearm development are homogeneous across subpopulations.  Among all orders in $O$, the 5-component GMR fitted with 6 covariates that have effect types determined by the forward procedure achieves the lowest BIC value and is therefore selected as the preferred model in terms of  covariate relevance and effect type.  The results of backward effect-type determination procedure are presented in Table \ref{BackwardDET}.  The backward effect-type determination procedure starts from a GMR model where all covariates have heterogeneous effect on the response.  Since all covariates have $p_0<0.05$ at step 1,  the procedure stops and  a 4-component GMR is preferred.  We compare the BIC values of the two preferred models and find that the 5-component GMR using the forward effect-type determination procedure has the lower BIC value (4137.43). It is therefore selected as the final model for fitting the bat data set. 

\begin{table}[h!] 
\centering 
  \caption { Forward procedure  for determining the effect type of 6 selected variables for the Chiroptera data.  } 
  \label{ForwardDET}
      \resizebox{\textwidth}{!}{
\begin{tabular}{cl  l l l l l  }  
\\[1mm]\hline 
\hline\\[-3mm]
\multicolumn{7}{l}{1-Het-Covariate GMR Model (heterogeneous: $X_1$,  homogeneous: others)}\\[1mm]
\hline\\[-3mm]
$X_1$&Precipitation&Temperature&$\log$(Population Density)&$\log$(Area)&$\log$(Body Mass)&AET\\[1mm]
$p_0$&$8.03*10^{-6}$ & $9.99*10^{-1}$ & $7.68*10^{-20}$ &
$3.74*10^{-2}$ & \bm{$6.27*10^{-59}$} & $7.90*10^{-18}$ \\[1mm]
\hline\\[-3mm]
\multicolumn{7}{l}{2-Het-Covariate GMR Model (heterogeneous: $\log(\text{Body Mass})$ + $X_2$,  homogeneous: others)}\\[1mm]
\hline\\[-3mm]
$X_2$&Precipitation&Temperature&$\log$(Population Density)&$\log$(Area)& AET&\\[1mm]
$p_0$&$1.87*10^{-2}$ & $5.79*10^{-4}$ & \bm{$3.52*10^{-4}$} &
$9.32*10^{-1}$ & $5.19*10^{-2}$  &\\[1mm]
\hline\\[-3mm]
\multicolumn{7}{l}{3-Het-Covariate GMR Model (heterogeneous: $\log(\text{Body Mass})$ + $\log$(Population Density)+ $X_3$,  homogeneous: others)}\\[1mm]
\hline\\[-3mm]
$X_3$&Precipitation&Temperature&$\log$(Area)& AET&&\\[1mm]
$p_0$&\bm{$4.29*10^{-5}$} & $8.16*10^{-4}$ & $6.25*10^{-1}$ & $2.27*10^{-3}$ &&\\[1mm]
\hline\\[-3mm]
\multicolumn{7}{l}{4-Het-Covariate GMR Model (heterogeneous: $\log(\text{Body Mass})$ + $\log$(Population Density)+ Precipitation + $X_4$,  homogeneous: others)}\\[1mm]
\hline\\[-3mm]
$X_4$&Temperature&$\log$(Area)& AET&&&\\[1mm]
$p_0$&$2.97*10^{-1}$ & $6.30*10^{-2}$ & $2.31*10^{-1}$  &&&\\[1mm]
\hline\\[-3mm]

Preferred model&\multicolumn{6}{l}{ \textbf{5-component GMR model with 6 covariates}}\\
&\multicolumn{6}{l}{Heterogeneous effect:  $\log(\text{Body Mass}),  \log(\text{Population Density}),  \text{Precipitation}$}\\
&\multicolumn{6}{l}{Homogeneous effect:  $\log(\text{Area)}, \text{AET}, \text{Temperature}$}\\
\hline
\hline
\end{tabular}}
\end{table} 

\begin{table}[h!] 
\centering 
  \caption {Backward  procedure for determining the effect type of 6 selected variables for the Chiroptera data.  } 
  \label{BackwardDET}
      \resizebox{\textwidth}{!}{
\begin{tabular}{c l  l l l l l }  
\\\hline 
\hline\\[-3mm]
\multicolumn{7}{l}{5-Het-Covariate GMR Model (homogeneous: $X_1$,  heterogeneous: others)}\\[1mm]
\hline \\[-3mm]
$X_1$&Precipitation&Temperature&$\log$(Population Density)&$\log$(Area)&$\log$(Body Mass)& AET\\[1mm]
$p_0$& $1.79*10^{-4}$ & $2.38*10^{-4}$ & $4.80*10^{-5}$ &
$2.49*10^{-3}$ & $1.49*10^{-28}$ & $2.97*10^{-5}$ \\[1mm]

\hline
\hline\\[-3mm]
Preferred model&\multicolumn{6}{l}{ \textbf{4-component GMR model with 6 covariates}}\\

&\multicolumn{6}{l}{Heterogeneous effect: $\log(\text{Body Mass}),  \log(\text{Population Density}), \log(\text{Area)}, \text{AET}, \text{Precipitation}, \text{Temperature}$ }\\[1mm]
\hline
\hline\\[-3mm]
\end{tabular}}
\end{table}

Table \ref{chiest2} lists parameter estimates of $\PI, \sigma$ and $\bt$ for the final model, and Figure S1(b) in Section 1 in the Supplemental Material presents the scatterplot of forearm length against log(Body Mass) and log(Population Density) based on the final model. The first two components shown in Table \ref{chiest2} are the major components where 40\% and 38\% bat species in the sample are assigned to these components respectively.  The covariate log (Body Mass) is the most significant factor of forearm length, showing positive coefficients in every mixture subpopulation.  Its influence, however, varies across subpopulation.  Two bats of equal mass can have different forearm lengths depending on wing loading, foraging strategy, habitat structure, and roosting constraints. For example,  the log(Body Mass) has the largest positive effect on the forearm length of bat species in the 4th component ($\beta_{14}=33.02$) which mainly contains herbivore bats.  $\log$(Population Density) is the second  covariate entering the model in the forward selection procedure and determined to have heterogeneous effects. It shows a strong negative effect on the forearm length in the third component, but in the other four components, it shows a positive effect with different magnitudes. These contrasting effects are consistent with patterns reported in the literature (\cite{ tomassini2014cranial, hughes2023big}).   Moderate urbanization often favors species with longer forearms, whereas intense urban development can eliminate them because their larger wingspans require more open space. The sizes of effect of Precipitation are relatively smaller in comparison to others, but yet remain significant.  The last three covariates, log(Area), AET and Temperature,  each is determined to have the same effect on forearm length across 5 subpopulations.

\begin{table}[!ht]
	\centering 
	\caption{Estimated parameters of 5-component GMR model  with 6 selected covariates that have effect types determined by the forward effect-type determination procedure (see Table \ref{ForwardDET}),  for the Chiroptera data.} 
	\label{chiest2}
	\begin{tabular}{cc c    } 
		\\[-2.8ex]\hline 
		\hline
		\textbf{Parameter}&\textbf{Covariate}&\textbf{Estimated parameter}\\
		\hline
        $\PI$ & & (0.38, 0.40, 0.09, 0.04, 0.09) \\
		$\sigma$ & & 4.03 \\
		$\bt_0$ & & (10.76, 10.07, 9.12, -55.38, 34.96) \\
		$\bt_1$ & $\log(\text{Body Mass})$ & (15.80, 12.25, 20.85, 33.02, 1.79) \\
		$\bt_2$ & $\log(\text{Population Density})$ & (0.03, 1.81, -2.69, 0.87, 1.46) \\
		$\bt_3$ & Precipitation & (0.06, 0.03, 0.11, 0.12, 0.04) \\
           \color{red}{$\bt_4$}&   \color{red}{$ \log(\text{Area)}$}&   \color{red}{(-0.28,   -0.28,  -0.28,   -0.28, -0.28)}\\
           \color{red}{$\bt_5$}&\color{red}{AET}&\color{red}{(-0.01,   -0.01,  -0.01,   -0.01,  -0.01)}\\
		\color{red}{$\bt_6$}&\color{red}{Temperature}&\color{red}{(0.03,   0.03,  0.03,  0.03,  0.03)}\\	

\hline
\hline
	\end{tabular} 
\end{table} 

\section{Simulation study and results}

To ascertain the performances of our proposed methods and to investigate their applicability, simulation studies are conducted. Our simulation studies  assess the performance of our proposed stepwise procedures for both variable selection and effect-type determination.  Key metrics for the assessment include recall and precision.  We also measure the accuracy of selecting relevant covariates and correct order of mixture model, as well as the adjusted Rand index to assess the precision of membership assignments, with detailed results reported in Section 2.2 in the Supplementary Material.  

\subsection{Simulation models}
The datasets are generated based on a 3-component GMR models with five covariates $\X=(\bx_1, \bx_2, \bx_3,  \bx_4, \bx_5)$.  We assume  5 covariates  are independent and each is generated from a standard normal distribution. A vector of binary variables $Z_i=(Z_{i1}, Z_{i2}, Z_{i3})^{'}$ is generated from a multinomial (1, $\PI$).  If $Z_{ig}=1$, the $i$th observation will be assigned to group $g$,  and $Y_i$ will be then generated from a Normal distribution with mean ($\beta_{0g}+\beta_{1g}x_{i1}+\beta_{2g}x_{i2}+\beta_{3g}x_{i3}+\beta_{4g}x_{i4}+\beta_{5g}x_{i5}$) and variance $\sigma^2$. Table \ref{settingSM1} lists the specified  parameter values of $\PI$'s, $\sigma$ and $\bt$'s used for simulating $Y$ given the simulated $\X's$ under two different scenarios.  In both scenarios, $\bx_1$ has heterogeneous and relatively strong  effects on $Y$; $\bx_2$ has a homogeneous and moderate effect on $Y$; $\bx_3$ has  heterogeneous but weak effects on $Y$; and $\bx_4$ and $\bx_5$ have  no effect on $Y$.
In Scenario 2, we halve the size of regression coefficients $\bt$ to reduce the covariate effects on the response for assessing their impact on recall and precision.  For each scenario,  a total of 1000 data sets are generated, with sample sizes of 500 and 300, respectively.

\begin{table}[!b]
\centering 
  \caption{True values of parameters in simulation study.} 
 \label{settingSM1}
\begin{tabular}{cc c    } 
\\[-2.8ex]\hline 
\hline
\textbf{Parameter}&\textbf{Scenario 1}&\textbf{Scenario 2}\\
\hline
$\PI$&(0.2, 0.3, 0.5)&(0.2, 0.3, 0.5)\\
$\sigma$&1&1\\
$\bt_{0}$&(-3, 0, 3)&(-3, 0, 3)\\
$\bt_{1}$& (0.3, 0.6, 0.8)& (0.15, 0.3, 0.4)\\
\color{red}{$\bt_{2}$}&\color{red}{(-0.4, -0.4, -0.4)}&\color{red}{(-0.2, -0.2, -0.2)}\\
$\bt_{3}$&(-0.2,  0,  0.3)&(-0.1,  0,  0.15)\\
\color{gray}{$\bt_{4}$}&\color{gray}{(0, 0, 0)}&\color{gray}{(0, 0, 0)}\\
\color{gray}{$\bt_{5}$}&\color{gray}{(0, 0, 0)}&\color{gray}{(0, 0, 0)}\\

\hline
\hline

\end{tabular} 
\end{table} 

Forward and backward procedures for variable selection and effect-type determination are applied sequentially to each simulated data set. After the procedures stops with a set of selected variable and their corresponding effect types,  the order will be selected according to the lowest BIC values among all order in $O$.   Since either a forward or backward procedure can be chosen at each of the two sequential steps, we consider all four possible combinations to produce the results.  We measure the accuracy of our stepwise procedures by using recall and precision.  Because variable selection is a binary task (relevant vs irrelevant), precision and recall can be applied directly to quantify how well the procedure identifies the truly relevant covariates. For each simulated data set, the results of of variable selection may differ, and therefore the set of covariates entering the effect-determination procedure is not fixed. In addition, the effect-type determination procedure is the final step of our sequential framework, classifying the effect type of each original covariate into one of three categories: heterogeneous, homogeneous, and no effect. Accordingly, we extend the definitions of precision and recall using a one-versus-rest formulation, in which  each effect type is treated as the ``positive” class in turn, with the remaining two types combined as the``negative” class. Precision and recall are then reported separately for each effect type.  For detailed explanations of the one-versus-rest precision and recall, please refer to Section 2.1 in the Supplementary Material.  Both metrics range from 0 to 1, with values closer to 1 indicating higher accuracy in classifying the  effect type.

\subsection{Simulation results}
For illustrative purposes, we only present simulation results using the forward variable selection procedure, followed by the forward or backward effect-type determination procedure.  A similar pattern can be observed using backward elimination procedure, as shown in Tables S10-S12 in Section 2.2 in the Supplementary Material. 

Table \ref{sim_TPVSF} summarizes the results obtained using the forward variable selection procedure. The precision and recall values are generally close to 1, indicating high accuracy in identifying relevant  covariates in the GMR model. Precision remains consistently high across both scenarios and sample sizes. However, recall is more sensitive to changes in sample size and the effect sizes of covariates. Specifically, when the sample size decreases from 500 to 300, the mean recall drops from 0.983 to 0.881 in Scenario 1, and from 0.686 to 0.501 in Scenario 2.  In Scenario 2,  $|\bt_1|$, $|\bt_2|$ and $|\bt_3|$ all decrease by half of the size. Recall is therefore expected to decline because covariates with weak effects  are harder to be detected.

In Tables \ref{sim_TPF} and \ref{sim_TPB}, the performance of the stepwise effect-type determination procedure as the final step of our sequential framework is evaluated using one-vesus-rest precision and recall across three effect types. As the effect size and sample size decrease, both the precision and recall decline for the heterogeneous and homogeneous effect types. However, the precision for heterogeneous effects decreases more gradually than the precision for homogeneous effects. This indicates a lower rate of falsely classifying a heterogeneous effect when the covariate truly has no effect or a homogeneous effect, compared with the false‑classification rate of labeling an effect as homogeneous when the covariate actually has no effect or a heterogeneous effect. Overall, the precisions of heterogeneous effects are higher than the precisions of  homogeneous effects. This suggests that among covariates determined as having heterogeneous effects, the proportion that truly have heterogeneous effects is generally higher than the proportion of truly homogeneous effects among covariates classified as having homogeneous effects. In contrast, the recalls of homogeneous effects are higher than those for heterogeneous effects. Since the true model contains only one covariate with a homogeneous effect, the high recall for the homogeneous effect indicates that  this covariate is correctly identified in a  large proportion of simulated datasets,  particularly when the sample size and effect size are large. By comparison, among covariates with truly heterogeneous effects, a smaller proportion are correctly classified as heterogeneous. This may reflect a trade-off between precision and recall.

\begin{table}[h!] 
	\small
	\centering 
	\caption{\footnotesize{Mean precision and recall using forward variable selection procedure.  Datasets were simulated under the settings in Table \ref{settingSM1}. }}
	\label{sim_TPVSF}

		\begin{tabular}{cc|c|c} 
			\\[-1.8ex]
			\hline
			\hline
			
			\textbf{Scenario } &&\multicolumn{1}{c|}{n=500}&\multicolumn{1}{c}{n=300}\\
			\hline

			1 & Precision&0.968 &0.959\\
			&Recall&0.983&0.881\\
			\hline
			\hline
		
			2 & Precision&0.955 &0.944\\
			&Recall&0.686&0.501\\
\hline
			\hline

	\end{tabular} 
\end{table} 

\begin{table}[h!] 
	\small
	\centering 
	\caption{\footnotesize{Mean precision and recall of the forward procedure for determining covariate effect types, applied after forward variable selection.}}
	\label{sim_TPF}

		\begin{tabular}{cc|ccc|ccc} 
			\\[-1.8ex]
			\hline
			\hline
			
			\textbf{Scenario } &&\multicolumn{3}{c|}{n=500}&\multicolumn{3}{c}{n=300}\\
			&Effect type& Het&Homo&No& Het&Homo&No\\
			\hline

			1 & Precision&0.946&0.883&0.982 &0.926&0.650&0.879\\
			&Recall&0.886&0.939&0.936&0.585&0.846&0.924\\
			\hline
			\hline

			2 & Precision&0.840&0.411&0.705 &0.798&0.268&0.586\\
			&Recall&0.302&0.591&0.936&0.185& 0.346&0.940\\
			\hline
			\hline

	\end{tabular} 
\end{table} 

\begin{table}[h!] 
	\small
	\centering 
	\caption{\footnotesize{Mean precision and recall of the backward procedure for determining covariate effect types, applied after forward variable selection.}}
	\label{sim_TPB}

		\begin{tabular}{cc|ccc|ccc} 
			\\[-1.8ex]
			\hline
			\hline
			
			\textbf{Scenario } &&\multicolumn{3}{c|}{n=500}&\multicolumn{3}{c}{n=300}\\
			&Effect type& Het&Homo&No& Het&Homo&No\\
			\hline

			1 & Precision&0.946&0.881&0.982 &0.928&0.654&0.879\\
			&Recall&0.885&0.938&0.936&0.590&0.846&0.924\\
			\hline
			\hline
2 & Precision&0.839&0.413&0.706 &0.794&0.260&0.597\\
			&Recall&0.305&0.592&0.936&0.187&0.362&0.926\\
			\hline
			\hline

	\end{tabular} 
\end{table}

\section{Conclusion and Discussion}
Mixture regression models are popular for clustering observations into  subpopulations when there are relationships between covariates and the response.  These models are especially useful when the data have underlying heterogeneity that reflects the different relationships between response and covariates across different subpopulations that standard regression models could not capture.  It has been applied in many real data analysis in various fields, but few studies make and report the inference decisions on the covariate effects on the response.  In the previous study,   \cite{gan2024tests} proposed  the testing procedures and provided a measure of statistical evidence for  the association between a covariate and the response under the null hypothesis of no association. In that study, only one covariate  was considered.  Real-world datasets often contain more than one covariate,  so our study expands the analysis to consider more covariates in the mixture regression model, and then propose the forward and backward variable selection and effect-type determination procedures.  Likelihood ratio test is a conventional statistical method to test the association between covariates and response.  However, in mixture regression framework, the order $G$ is often not pre-specified,  and the best fitted model among orders in the candidate list $O$ under $H_0$  might select an optimal order different than the one under $H_a$.  WEST procedure overcomes this issue by aggregating  statistical evidence across all candidate orders through a weighted significance level $p_0$. Consequently, the WEST is applicable in both the stepwise variable selection and effect‑type determination procedures, for both the forward and backward options. Through the WEST procedure,  we can use $p_0$ to quantify the significance of the covariate effect in each variable selection and effect-type step and identify a parsimonious model that best fits the data. However, our method does not provide the significance level for each covariate selected in the final model regardless their effect types are classified as homogeneous or heterogeneous. Since the data are used for variable selection, effect‑type determination, and model selection, conducting post‑selection inference, such as hypothesis testing, on these same data may lead to biased results. When the sample size is sufficiently large, a data‑splitting approach can be used, in which the data are split into a selection set for model selection and an inference set for parameter estimation and hypothesis testing in the selected model.

Bats represent one of the largest groups of mammals, comprising more than 20\% of extant species and standing as the only mammals capable of true flight. We apply our method to Chiroptera data to investigate how natural selection has shaped bat forearm development in response to dynamic environmental and biogeographical pressures. It is also important to examine how the evolution of forearm length relates to phylogenetic patterns revealed by molecular data  (\cite{teeling2005molecular, ushiki2026functional}). In particular, incorporating phylogenetic information from key genes involved in bat forelimb development, such as \textit{Prx1} (\cite{cretekos2008regulatory}) and \textit{Hoxd10–Hoxd13} (\cite{chen2005hoxd13, booker2016bat}), into the mixture regression model may help elucidate their roles while accounting for diverse environmental influences. Possible ways to incorporate phylogenetic information include using phylogenetic eigenvectors derived from principal coordinates analysis of the sequencing data, as well as incorporating evolutionary molecular rates estimated from nucleotide substitution patterns in these genes. Our proposed method provides a foundation for pursuing this direction further in future research.

Finally,  our proposed stepwise variable selection procedures only retain influential covariates and remove irrelevant covariates by setting their coefficients to be all zero across all subpopulations. In this sense, our proposed method can be considered as a groupwise selection method. While this groupwise approach is effective for detecting overall variable relevance, it does not select variables that have effects within individual subpopulations specifically. Inspired by the work of \cite{Khalili2007}, who introduced regularized penalty functions  at the individual $\beta_{jg}$ in finite mixture regression models for component-specific variable selection, it would be worthwhile to develop complementary procedures to test whether a covariate in a given component has no effect, that is to test whether $\beta_{jg} = 0$ or not, while allows the covariate has or doesn't have an effect on the response in other components.  Moreover, the current methods are developed  for continuous response variables through the Gaussian mixture regression model.  Our methods can be extended to other distributions within the exponential family such as Poisson, Binomial and Gamma distributions. These extensions would broaden the applicability of the method to count data and binary outcomes, which are common in ecological, biomedical, and social science research.


\begin{funding}
This work was supported by Discovery Grant from Natural Sciences and Engineering Research Council of Canada  (grant number 400095 to ZF). 
\end{funding}

\begin{supplement}
\stitle{1. Additional information on Chiroptera data}
\sdescription{Table S1, Figure S1}
\end{supplement}
\begin{supplement}
\stitle{2. Simulation studies for sequential stepwise procedures}
\sdescription{[2.1] Recall and precision (Tables S2-S3); [2.2] Additional simulation results (Tables S4-S12)}
\end{supplement}
\begin{supplement}
\stitle{3. Source code}
\end{supplement}


\bibliographystyle{imsart-nameyear} 
\bibliography{biblio}       


\end{document}


\articletype{}

\title{Supplementary Materials for \\ ``Hypothesis Test Based Stepwise Model Selection in Finite Gaussian Mixture Regression Models"}


\author{
\name{Chong Gan\textsuperscript{a}, Jacqueline May\textsuperscript{b} and Zeny Feng\textsuperscript{c}\thanks{CONTACT: Zeny Feng. Email: zfeng@uoguelph.ca}}
\affil{\textsuperscript{a}Department of Public Health Sciences, Queen's University; \textsuperscript{b}School of Pharmacy,  University of Waterloo; \textsuperscript{c}Department of Mathematics and Statistics, University of Guelph  }
}

\maketitle
\section{Additional information on Chiroptera data}
\label{app:C}
\begin{figure}[!h]
    \centering
    
    \begin{minipage}[b]{0.54\textwidth}

    \begin{overpic}[width=\textwidth]{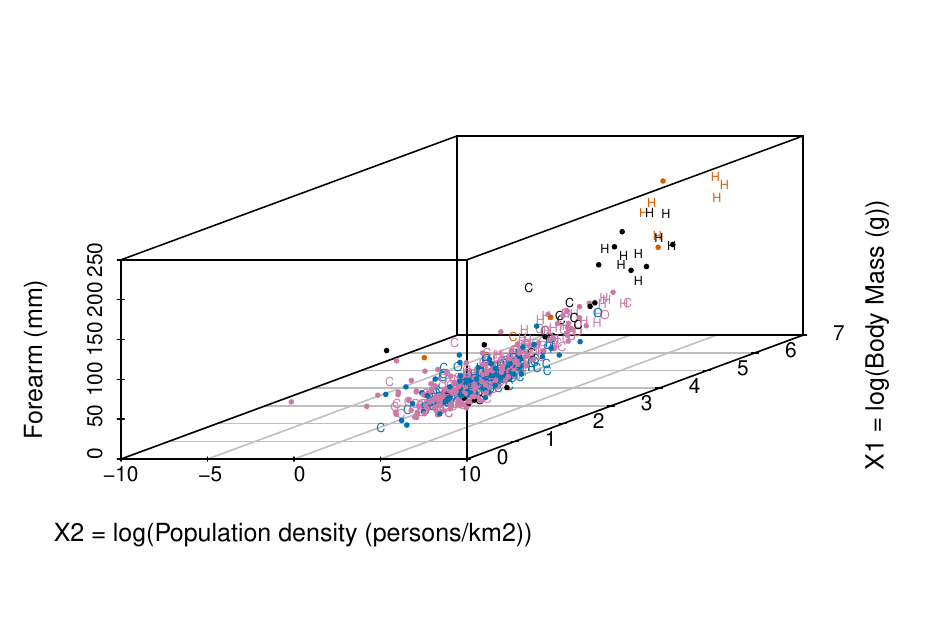}
            \put(50,2){\makebox(0,0){\textcolor{black}{(a) 4-GMR with 6 covariates}}}
        \end{overpic}
    \end{minipage}
          \begin{minipage}[b]{0.54\textwidth}

    \begin{overpic}[width=\textwidth]{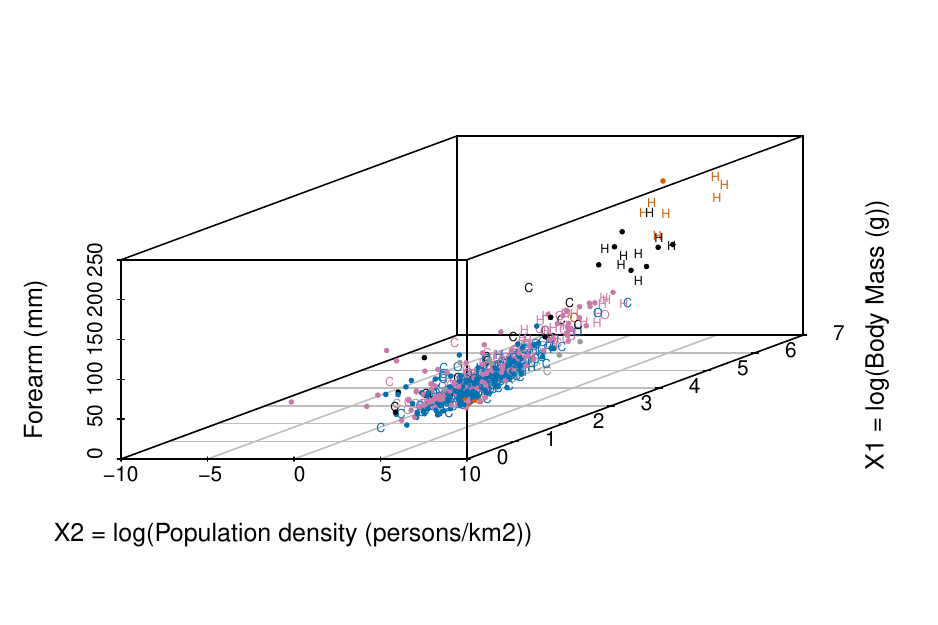}
            \put(50,2){\makebox(0,0){\textcolor{black}{(b) 5-Homo-GMR with  6 covariates}}}
        \end{overpic}
    \end{minipage}
    \caption{ The scatterplots of Chiroptera data,  using (a) 4-component GMR model with 6 covariates selected by backward effect-type determination procedure, and (b) 5-component Homo-GMR model with 6 selected covariates that have effect types determined by the forward effect-type determination procedure,  where colors refer to the fitted groups, letters refer to the classified groups: H (herbivore), C (carnivore), and O (omnivore), and unmarked dots refer to the species that lack information on trophic level.}
    \label{A:fig:bat1}
\end{figure}

\begin{table}[!h]
\centering 
  \caption{Estimated parameters of 4-component GMR model with 6 covariates as in Table 4 for the Chiroptera data. } 
 \label{chiest2}
\begin{tabular}{cc c    } 
\\[-2.8ex]\hline 
\hline
\textbf{Parameter}&\textbf{Covariate}&\textbf{Estimate}\\
\hline
$\PI$      &               & (0.48, 0.36, 0.13, 0.04) \\
$\sigma$   &               & 4.51                     \\
$\bt_0$    &               & (9.12, 22.75, 6.61, -168.62) \\
$\bt_1$    & $\log(\text{Body Mass})$         & (14.91, 8.41, 26.90, 27.47)   \\
$\bt_2$    & $\log(\text{Population Density})$ & (0.30, 1.43, 0.15, -0.14)     \\
$\bt_3$    & $\log(\text{Area})$               & (-0.24, -0.33, -1.11, 3.24)   \\
$\bt_4$    & AET                               & (-0.01, -0.00, -0.04, -0.05)  \\
$\bt_5$    & Precipitation                     & (0.05, 0.00, 0.18, 0.39)      \\
$\bt_6$    & Temperature                       & (0.04, -0.01, 0.02, 0.48)     \\

\hline
\hline
\end{tabular} 
\end{table}

\section{Simulation studies for sequential stepwise procedures}
\label{app:B}
\subsection{Recall and Precision}
\label{app:B1}
Recall and precision are used to assess the accuracy of our proposed stepwise procedures, and they depend on the counts of true positive (TP), false positive (FP), true negative (TN) and false negative (FN).  Specifically,  $recall=\frac{TP}{TP+FN}$ and $precision=\frac{TP}{TP+FP}$.  

When we measure the variable selection accuracy alone, recall and precision are computed based on the $2\times 2$ confusion matrix shown in Table  \ref{Matrix1}, where $TP=a_{11}$, $FP=a_{21}$, $FN=a_{12}$ and $TN=a_{22}$.  For example, if a variable truly has an effect on the response and is selected as a relevant covariate by the stepwise variable selection procedure, it is classified as TP. Conversely, if it is incorrectly identified as irrelevant, it is classified as FN.

\begin{table}[h!]
\caption{Confusion matrix used to evaluate variable selection performance. Each variable either truly has an effect or does not, and is either estimated as having an effect or not. }
\label{Matrix1}
\centering
\begin{tabular}{c|cc}
\hline
\multirow{2}{*}{\textbf{True Effect}} & \multicolumn{2}{c}{\textbf{Estimated Effect}} \\
& Have effect & No effect \\
\hline
Have effect & $a_{11}$ & $a_{12}$ \\
No effect     & $a_{21}$ & $a_{22}$ \\
\hline
\end{tabular}
\end{table}

\begin{table}[h!]
\caption{Confusion matrix used to evaluate the overall performance of sequential stepwise procedures.  Each variable is classified based on its true effect type and its estimated effect type (i.e. heterogeneous, homogeneous, or null).}
\label{Matrix2}
\centering
\begin{tabular}{c|ccc}
\hline
\multirow{2}{*}{\textbf{True Effect Type}} & \multicolumn{3}{c}{\textbf{Estimated Effect Type}} \\
& Heterogeneous & Homogeneous & No effect\\
\hline
Heterogeneous & $a_{11}$ & $a_{12}$& $a_{13}$ \\
Homogeneous & $a_{21}$ & $a_{22}$& $a_{23}$ \\
No  effect& $a_{31}$ & $a_{32}$&$a_{33}$ \\
\hline
\end{tabular}
\end{table}

When we measure the overall performance of sequential stepwise procedures,  recall and precision are computed based on the $3\times 3$ confusion matrix shown in Table  \ref{Matrix2}. Since now the effect types of variables have three classes (heterogeneous, homogeneous,  and no effect), the precision and recall are defined for each class respectively. 
For example,  in  the heterogeneous class, if a variable truly has a heterogeneous effect on the response and is correctly identified as such during the stepwise procedure, it is classified as TP,  while if the variable is incorrectly identified as having a homogeneous effect or no effect, it is classified as FN.  Thereby, for each class $k \in \{1,  2,  3\}$, $recall_k=\frac{a_{kk}}{\sum_{i=1}^3a_{ik}}$ and $precision_k=\frac{a_{kk}}{\sum_{j=1}^3a_{kj}}$.
\subsection{Simulation results}
\label{app:B2}

\begin{table}[h!] 
	\small
	\centering 
	\caption{\footnotesize{Rank of covariates entering into the model,  selection rate of each covariates and the percentage of correctly selected all relevant covariates.  Datasets were simulated under the settings in Table 6. The forward variable selection procedure was conducted.}}
	\label{sim_fs}
	\resizebox{\textwidth}{!}{
		\begin{tabular}{cc|c|c} 
			\\[-1.8ex]
			\hline
			\hline
			
			\textbf{Scenario } &&\multicolumn{1}{c|}{n=500}&\multicolumn{1}{c}{n=300}\\
			\hline

			1&Rank of entering into the model ($\bx_1,  \bx_2,  \bx_3,  \bx_4,  \bx_5$)&1.006,  2.185, 2.810,  3.862, 3.906& 1.046, 2.166, 2.686, 3.573, 3.688\\
			&Selection rate ($\bx_1,  \bx_2,  \bx_3,  \bx_4, \bx_5$)& 1, 0.997, 0.951, 0.065, 0.064&1, 0.905, 0.739, 0.075,  0.077 \\
             & \% of correct covariates selected&0.826&0.578\\
			\hline
			\hline
			
			2&Rank of entering into the model ($\bx_1,  \bx_2,  \bx_3,  \bx_4,  \bx_5$)&1.084,  2.052, 2.425,  2.804, 2.808& 1.120, 1.891,  2.056,  2.226, 2.322\\
			&Selection rate ($\bx_1,  \bx_2,  \bx_3,  \bx_4, \bx_5$)& 0.986, 0.655, 0.416, 0.056, 0.073&0.859, 0.394, 0.250, 0.062,  0.059 \\
             & \% of correct covariates selected&0.246&0.083\\
			\hline
			\hline

	\end{tabular} }
\end{table} 

\begin{table}[h!] 
	\small
	\centering 
	\caption{\footnotesize{Heterogeneity selection rate,  mean ARI, percentage of correctly determined covariates' effect types, orders and models. Datasets were simulated under the settings in Table 6. The forward procedure on determining covariates' effect types was conducted after the forward variable selection.}}
	\label{sim_fsf}
	\resizebox{\textwidth}{!}{
		\begin{tabular}{cc|c|c} 
			\\[-1.8ex]
			\hline
			\hline
			
			\textbf{Scenario } &&\multicolumn{1}{c|}{n=500}&\multicolumn{1}{c}{n=300}\\
			\hline

			1&Heterogeneity selection rate ($\bx_1,  \bx_2,  \bx_3,  \bx_4,  \bx_5$)&0.896, 0.058, 0.875, 0.038, 0.046& 0.599, 0.059, 0.570, 0.038, 0.034\\
						&Homogeneity selection rate ($\bx_1,  \bx_2,  \bx_3,  \bx_4,  \bx_5$)&0.104, 0.939, 0.076, 0.027, 0.018& 0.401, 0.846, 0.169, 0.037, 0.043\\
			             &Mean ARI&0.746&0.687\\
             & \% of correct effect types for all $\bx$s&0.654&0.257\\
             & \% of correct orders&0.977&0.789\\
             & \% of correct models&0.646&0.225\\

			\hline
			\hline
		
			2&Heterogeneity selection rate ($\bx_1,  \bx_2,  \bx_3,  \bx_4,  \bx_5$)&0.321, 0.064, 0.282, 0.029, 0.041& 0.212, 0.050, 0.157, 0.030, 0.026\\
						&Homogeneity selection rate ($\bx_1,  \bx_2,  \bx_3,  \bx_4,  \bx_5$)&0.665, 0.591, 0.134, 0.027, 0.032& 0.650, 0.346, 0.094,  0.032,  0.033\\
			             &Mean ARI&0.750&0.707\\
             & \% of correct effect types for all $\bx$s&0.054&0.011\\
             & \% of correct orders&0.977&0.832\\
             & \% of correct models&0.052&0.011\\
			\hline
			\hline

	\end{tabular} }
\end{table} 

\begin{table}[h!] 
	\small
	\centering 
	\caption{\footnotesize{Heterogeneity selection rate,  mean ARI, percentage of correctly determined covariates' effect types, orders and models. Datasets were simulated under the settings in Table 6. The backward procedure on determining covariates' effect types was conducted after the forward variable selection.}}
	\label{sim_fsB}
	\resizebox{\textwidth}{!}{
		\begin{tabular}{cc|c|c} 
			\\[-1.8ex]
			\hline
			\hline
			
			\textbf{Scenario } &&\multicolumn{1}{c|}{n=500}&\multicolumn{1}{c}{n=300}\\
			\hline

			1&Heterogeneity selection rate ($\bx_1,  \bx_2,  \bx_3,  \bx_4,  \bx_5$)&0.895, 0.059, 0.875, 0.038, 0.046& 0.605, 0.059, 0.574, 0.037, 0.034\\
			&Homogeneity selection rate ($\bx_1,  \bx_2,  \bx_3,  \bx_4,  \bx_5$)&0.105, 0.938, 0.076, 0.027, 0.018& 0.395, 0.846, 0.165, 0.038, 0.043 \\
			             &Mean ARI&0.746&0.686\\
             & \% of correct effect types for all $\bx$s &0.652&0.263\\
             & \% of correct orders&0.976&0.786\\
             & \% of correct models&0.644&0.228\\

			\hline
			\hline
			
			2&Heterogeneity selection rate ($\bx_1,  \bx_2,  \bx_3,  \bx_4,  \bx_5$)&0.326, 0.065, 0.284, 0.030, 0.042& 0.215, 0.049, 0.158,  0.034, 0.030\\
			&Homogeneity selection rate ($\bx_1,  \bx_2,  \bx_3,  \bx_4,  \bx_5$)&0.662, 0.592, 0.133, 0.026, 0.031& 0.704, 0.362, 0.108, 0.034, 0.030 \\
			             &Mean ARI&0.751&0.706\\
             & \% of correct effect types for all $\bx$s &0.055&0.012\\
             & \% of correct orders&0.976&0.828\\
             & \% of correct models&0.053&0.012\\

			\hline
			\hline

	\end{tabular} }
\end{table} 

\begin{table}[h!] 
	\small
	\centering 
	\caption{\footnotesize{Rank of covariates removing from the model,  selection rate of each covariates and the percentage of correctly selected all relevant covariates.  Datasets were simulated under the settings in Table 6. The backward variable elimination procedure was conducted.}}
	\label{sim_bs}
	\resizebox{\textwidth}{!}{
		\begin{tabular}{cc|c|c} 
			\\[-1.8ex]
			\hline
			\hline
			
			\textbf{Scenario } &&\multicolumn{1}{c|}{n=500}&\multicolumn{1}{c}{n=300}\\
			\hline

			1&Rank of removing from the model ($\bx_1,  \bx_2,  \bx_3,  \bx_4,  \bx_5$)&2.000,  2.250,  2.648,  1.509,  1.450& 1.417, 2.615 2.498, 1.568,  1.586\\
			&Selection rate ($\bx_1,  \bx_2,  \bx_3,  \bx_4, \bx_5$)& 0.998, 0.996, 0.946, 0.067, 0.059&0.988,  0.891,  0.721,  0.077,  0.071 \\
             & \% of correct covariates selected&0.820&0.552\\
			\hline
			\hline
			
			2&Rank of removing from the model($\bx_1,  \bx_2,  \bx_3,  \bx_4,  \bx_5$)&2.241,  2.784,  2.423,  1.761,  1.766& 3.189, 2.767,  2.511,  2.090,  1.984\\
			&Selection rate ($\bx_1,  \bx_2,  \bx_3,  \bx_4, \bx_5$)& 0.971, 0.635, 0.408, 0.059, 0.071&0.825, 0.369, 0.245, 0.065,  0.057 \\
             & \% of correct covariates selected&0.238&0.072\\
			\hline
			\hline

	\end{tabular} }
\end{table}

\begin{table}[h!] 
	\small
	\centering 
	\caption{\footnotesize{Heterogeneity selection rate,  mean ARI, percentage of correctly determined covariates' effect types, orders and models. Datasets were simulated under the settings in Table 6. The forward procedure on determining covariates' effect types was conducted after the backward variable elimination.}}
	\label{sim_Bsf}
	\resizebox{\textwidth}{!}{
		\begin{tabular}{cc|c|c} 
			\\[-1.8ex]
			\hline
			\hline
			
			\textbf{Scenario } &&\multicolumn{1}{c|}{n=500}&\multicolumn{1}{c}{n=300}\\
			\hline

			1&Heterogeneity selection rate ($\bx_1,  \bx_2,  \bx_3,  \bx_4,  \bx_5$)&0.891, 0.059, 0.868, 0.037, 0.040& 0.592, 0.058, 0.546, 0.036, 0.027\\
						&Homogeneity selection rate ($\bx_1,  \bx_2,  \bx_3,  \bx_4,  \bx_5$)&0.107, 0.937, 0.078, 0.030, 0.019& 0.396, 0.833, 0.175, 0.041, 0.044\\
			             &Mean ARI&0.744&0.685\\
             & \% of correct effect types for all $\bx$s&0.649&0.246\\
             & \% of correct orders&0.967&0.784\\
             & \% of correct models&0.638&0.215\\

			\hline
			\hline
		
			2&Heterogeneity selection rate ($\bx_1,  \bx_2,  \bx_3,  \bx_4,  \bx_5$)&0.316,  0.065, 0.270, 0.032, 0.043& 0.201, 0.044, 0.153, 0.029, 0.025\\
						&Homogeneity selection rate ($\bx_1,  \bx_2,  \bx_3,  \bx_4,  \bx_5$)&0.653, 0.569, 0.138, 0.027, 0.028& 0.624, 0.325, 0.092,  0.036, 0.032\\
			             &Mean ARI&0.748&0.700\\
             & \% of correct effect types for all $\bx$s&0.053&0.010\\
             & \% of correct orders&0.964&0.807\\
             & \% of correct models&0.051&0.009\\
			\hline
			\hline

	\end{tabular} }
\end{table} 

\begin{table}[h!] 
	\small
	\centering 
	\caption{\footnotesize{Heterogeneity selection rate,  mean ARI, percentage of correctly determined covariates' effect types, orders and models. Datasets were simulated under the settings in Table 6. The backward procedure on determining covariates' effect types was conducted after the backward variable elimination.}}
	\label{sim_BsB}
	\resizebox{\textwidth}{!}{
		\begin{tabular}{cc|c|c} 
			\\[-1.8ex]
			\hline
			\hline
			
			\textbf{Scenario } &&\multicolumn{1}{c|}{n=500}&\multicolumn{1}{c}{n=300}\\
			\hline

			1&Heterogeneity selection rate ($\bx_1,  \bx_2,  \bx_3,  \bx_4,  \bx_5$)&0.889, 0.060, 0.868, 0.041, 0.042& 0.593, 0.057, 0.551, 0.045,  0.034\\
			&Homogeneity selection rate ($\bx_1,  \bx_2,  \bx_3,  \bx_4,  \bx_5$)&0.109, 0.936, 0.078, 0.026, 0.017& 0.395, 0.834,  0.170, 0.032, 0.037 \\
			             &Mean ARI&0.744&0.684\\
             & \% of correct effect types for all $\bx$s &0.646&0.251\\
             & \% of correct orders&0.972&0.779\\
             & \% of correct models&0.636&0.217\\

			\hline
			\hline
			
			2&Heterogeneity selection rate ($\bx_1,  \bx_2,  \bx_3,  \bx_4,  \bx_5$)&0.324, 0.065, 0.273, 0.034, 0.045& 0.207, 0.045, 0.156,  0.036,  0.031\\
			&Homogeneity selection rate ($\bx_1,  \bx_2,  \bx_3,  \bx_4,  \bx_5$)&0.647, 0.570, 0.135, 0.025,  0.026& 0.618, 0.324, 0.089,  0.029, 0.026 \\
			             &Mean ARI&0.749&0.700\\
             & \% of correct effect types for all $\bx$s &0.054&0.010\\
             & \% of correct orders&0.974&0.814\\
             & \% of correct models&0.052&0.010\\

			\hline
			\hline

	\end{tabular} }
\end{table} 

\begin{table}[h!] 
	\small
	\centering 
	\caption{\footnotesize{Mean precision and recall using backward variable elimination procedure.  Datasets were simulated under the settings in Table 6. }}
	\label{sim_TPVSB}

		\begin{tabular}{cc|c|c} 
			\\[-1.8ex]
			\hline
			\hline
			
			\textbf{Scenario } &&\multicolumn{1}{c|}{n=500}&\multicolumn{1}{c}{n=300}\\
			\hline

			1 & Precision&0.969 &0.960\\
			&Recall&0.980&0.867\\
			\hline
			\hline
		
			2 & Precision&0.954 &0.943\\
			&Recall&0.671&0.480\\
\hline
			\hline

	\end{tabular} 
\end{table} 

\begin{table}[h!] 
	\small
	\centering 
	\caption{\footnotesize{Mean precision and recall of the forward procedure for determining covariate effect types, applied after backward variable elimination procedure.}}
	\label{sim_TPBF}

		\begin{tabular}{cc|ccc|ccc} 
			\\[-1.8ex]
			\hline
			\hline
			
			\textbf{Scenario } &&\multicolumn{3}{c|}{n=500}&\multicolumn{3}{c}{n=300}\\
			&Effect type& Het&Homo&No& Het&Homo&No\\
			\hline

			1 & Precision&0.949&0.879&0.979 &0.930&0.643&0.866\\
			&Recall&0.880&0.937&0.937&0.569&0.833&0.926\\
			\hline
			\hline

			2 & Precision&0.829&0.403&0.696 &0.805&0.265&0.575\\
			&Recall&0. 293&0.569&0.935&0.177& 0.325&0.939\\
			\hline
			\hline

	\end{tabular} 
\end{table} 

\begin{table}[h!] 
	\small
	\centering 
	\caption{\footnotesize{Mean precision and recall of the backward procedure for determining covariate effect types, applied after backward variable elimination procedure.}}
	\label{sim_TPBB}

		\begin{tabular}{cc|ccc|ccc} 
			\\[-1.8ex]
			\hline
			\hline
			
			\textbf{Scenario } &&\multicolumn{3}{c|}{n=500}&\multicolumn{3}{c}{n=300}\\
			&Effect type& Het&Homo&No& Het&Homo&No\\
			\hline

			1 & Precision&0.947&0.878&0.979 &0.927&0.649&0.865\\
			&Recall&0.879&0.936&0.937&0.572&0.834&0.926\\
			\hline
			\hline
2 & Precision&0.832&0.407&0.690 &0.794&0.269&0.575\\
			&Recall&0.299&0.570&0.935&0.182&0.364&0.939\\
			\hline
			\hline

	\end{tabular} 
\end{table} 

For forward variable selection procedure,  we assess the performance of identifying influential covariates by computing the mean rank of each covariate entering into the model and the selection rate for each covariate over 1000 replications.  A lower rank value for a covariate indicates that it is added to the model earlier,  suggesting its effect is more significant than that of the others.  On the other hand, the selection rate of each covariate reflects its consistency being selected by the procedure.  For backward elimination procedure,  the rank of removal for a covariate measures its order of being removing from the model if it is removed before the procedure stop. A lower rank value for a covariate suggests that the variable holds less significance.   Finally, the mean adjusted Rand index (ARI) and  the proportions of correctly selected order, effect types, and model are measured to evaluate the clustering performance for the  preferred model after the sequential stepwise procedures.

\section{Source code}
We used R to implement all the methods.  The dataset and source code,  including functions for stepwise variable selection and effect-type determination,  are available at \url{https://github.com/c2gan/MS-GMR}.
\label{app:EM}


 